\documentclass[journal]{IEEEtran}

\usepackage[noadjust]{cite}
\usepackage{datetime}
\usepackage{comment}
\usepackage{amsmath}
\usepackage{amssymb}
\usepackage{hyperref}
\usepackage{color}
\DeclareMathOperator*{\argmax}{argmax}

\usepackage{graphicx}
\begin{document}
	
\title{Taming Spontaneous Stop-and-Go Traffic Waves:\\ A Computational Mechanism Design Perspective}
\author{Di Shen, \;Qi Dai, \;Suzhou Huang*,\; Dimitar Filev*\\
\thanks{D. Shen, Q. Dai, S. Huang are independent researchers (e-mail: \{saddieyoyo, daiqi5477, huang0suzhou\}@gmail.com)}
\thanks{D. Filev is with Texas A\&M University (e-mail: filev@tamu.edu)}
\thanks{* Corresponding authors}
}

\markboth{}%
{Taming waves}
  
\maketitle

\begin{abstract}
It is well known that stop-and-go waves can be generated spontaneously in traffic even without bottlenecks. Can such undesirable traffic patterns, induced by intrinsic human driving behaviors, be tamed effectively and inexpensively? Taking advantage of emerging connectivity and autonomy technologies, we envision a simple yet realistic traffic control system to achieve this goal. To prove the concept, we design such a system to suppress these waves while maximizing traffic throughput in the Tadaki setting: a circular road with varying number of vehicles. We first introduce our driver behavior model and demonstrate how our calibrated human driving agents can closely reproduce the observed human driving patterns in the original Tadaki experiment. We then propose a simple control system mediated via connected automated vehicles (CAV) whose ideal speed parameter is treated as a system-level control variable adapted to the local vehicle density of the traffic. The objective of the control system is set up as a tradeoff: maximizing throughput while minimizing traffic oscillation. Following computational mechanism design, we search for the optimal control policy as a function of vehicle density and the tradeoff attitude parameter. This can be done by letting all vehicles play a simulated game of CAV-modulated traffic under such a control system. Our simulation results show that the improvements in traffic efficiency and smoothness are substantial. Finally, we envision how such a traffic control system can be realized in an environment with smart vehicles connected to a smart infrastructure or via a scheme of variable speed advisory.
\end{abstract}

\begin{IEEEkeywords}
	taming traffic waves, calibrated human driving agents, computational mechanism design
\end{IEEEkeywords}	

\IEEEpeerreviewmaketitle

\section{\bf Introduction}
\begin{figure}[!h]
	\centering
	\includegraphics[width=0.46\textwidth]{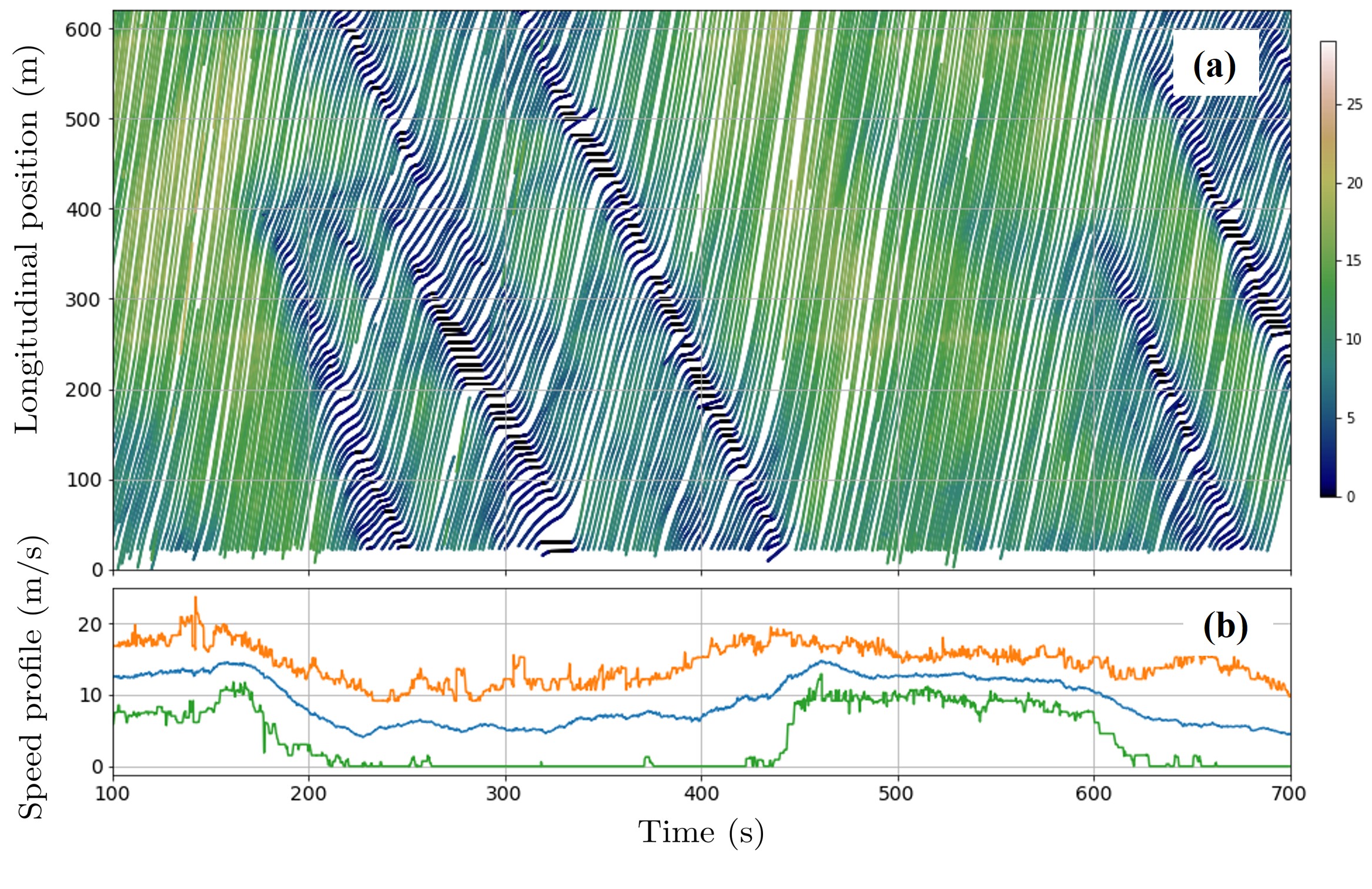}
	\caption{(a) Vehicle trajectories of the innermost lane on a stretch of US101 in California derived from a naturalistic traffic video recorded by NGSIM. The color legend represents instantaneous speed of the vehicle in m/s. Stop-and-go waves are clearly identifiable; (b) Aggregate speed profiles: average (blue), minimum (green) and maximum (orange).}
	\label{US101_waves}
\end{figure}

The phenomena of stop-and-go wave formation are commonly observed in a variety of traffic settings. Of course, traffic oscillations can be induced by external disturbances, such as vehicle merges/lane-changes, road bottlenecks or road condition changes. However, as demonstrated by the Sugiyama experiment \cite{Sugiyama08}, such waves can also be generated spontaneously even without any traffic bottlenecks, purely due to human driving behaviors. The Tadaki experiment \cite{Tadaki13} further showed that these types of traffic oscillations have a non-trivial dependence on vehicle density. In a more naturalistic setting, similar waves are clearly identified from vehicle trajectories derived from traffic videos recorded by \cite{NGSIMdata} on a stretch of highway in California (US101 with multiple lanes), shown in Fig.\ref{US101_waves}. A closer examination of these trajectories seems to indicate that these waves are mostly formed by vehicles within the same lane, though they may be triggered by external disturbances, such as sudden vehicle lane changes. The figure also shows the aggregate speed profiles, consisting of maximum, minimum and average speeds, at each time step. The waves are best signaled by a very low minimum-speed and a very high speed-range, though average speed and vehicle density can also be good indicators. Left unchecked, such waves hinder traffic flow efficiency and burden the vehicles involved with excessive braking and acceleration, which in turn exacerbates fuel consumption and polluting emissions, and also extra wear and tear. 

There is a huge body of literature that attempts to understand traffic flow from the microscopic and macroscopic perspectives in general and the origins of the stop-and-go wave formation in particular. It is well beyond the scope of this paper to review them systematically. We refer the following books, \cite{Treiber2013, Garavello2016, Kerner2017}, for the latest developments using a variety of approaches. 

Recent studies collectively enhanced our understanding of traffic patterns and model accuracy, emphasizing the significance of human factors in traffic models. \cite{Bando1998}  \cite{Davis2003}  \cite{Nakayama2016} investigate the optimal velocity (OV) model, highlighting its role in predicting and improving traffic flow and jamming phenomena. \cite{Treiber2017} examines the impact of random fluctuations on traffic flow and stability using the Intelligent Driver Model (IDM) under stochastic conditions. \cite{Bando1995a} \cite{Swaroop1996}  \cite{Orosz2006} analyze dynamical models of traffic congestion and string stability in interconnected systems.

Early research on taming traffic jams has largely focused on optimizing traffic signal control and considering scenarios with complex traffic networks such as merges and intersections using traffic flow models \cite{Michalopoulos1981AnAO}. \cite{Eddelbttel1994ANA} derived an optimal control policy for traffic lights, taking into account queue length models and constraints such as capacity, and tested it in simulations. \cite{Lo1999ANT} \cite{Lo2001ACT} modeled traffic using the cell-transmission model (CTM) and proposed dynamic timing plans adaptable to various traffic conditions. In severely congested traffic, \cite{Michalopoulus1988ANALYSISOT} proposed practical solutions utilizing both sign and signal control. Similarly, studies have coordinated traffic flow models with optimal control strategies like Model Predictive Control (MPC) to guide conventional traffic control systems such as ramp metering\cite{Stephanedes1993OptimalCO} \cite{Papageorgiou2002}, VSA (variable speed advisory) or VSL (variable speed limit)  \cite{Hegyi2003MPCbasedOC} \cite{Breton2002ShockWE} or both \cite{Han2017} \cite{Hegyi2002OptimalCO} to alleviate congestion at freeway bottlenecks or ramp road sections. It is noteworthy that among the literature reviewed, most focus on reducing or eliminating shock waves, with only \cite{Hegyi2003MPCbasedOC} addressing the simultaneous impact on reducing congestion, increasing outflow, and lowering the total time spent. On the other hand, signal control research has predominantly been from a macro-level perspective, overlooking the effect of spontaneous human driver responses.

Adaptive Cruise Control (ACC) and Cooperative Adaptive Cruise Control (CACC), on the other hand, study simpler traffic scenarios such as car-following where there is limited traffic disturbance. The impact of ACC systems on traffic flow has been analyzed using the Intelligent Driver Model (IDM) in extensive simulations. While most findings suggest that ACC can increase highway capacity, they also indicate that it may exacerbate jams \cite{Davis2004}, \cite{JerathAPS2012}. CACC, however, appears more promising in increasing lane capacity and energy efficiency, especially at higher market penetrations \cite{Milanes2014a}, \cite{Milanes2014b}, \cite{Shladover2012ImpactsoC}, \cite{GE2018}, though the necessity of explicit coordination and communication between vehicles (V2V) falls outside the scope of this study. Arguably, the IDM, as a model of individual car behavior, may not adequately represent collective traffic phenomena \cite{CalibrationPaper}, making it less persuasive to conclude a positive real-life implementation effect based solely on simulations.

The emergence of Autonomous Vehicles (AV) and Connected Autonomous Vehicles (CAV) has significantly enhanced the adaptability of traffic management. A study by \cite{Lord2022} comprehensively investigates various vehicle dynamic models in stop-and-go traffic, utilizing the Flow platform for extensive simulations. It assessed key metrics such as AV penetration rates, stabilization time, headway, vehicle miles traveled, and fuel economy. The findings underscore that reinforcement learning (RL) controllers consistently deliver improvements across various conditions. Likewise, research employing Deep-RL \cite{Freidieh2018} demonstrates effective traffic control strategies and flow improvements with low AV penetration. The CIRCLES Consortium's experiment \cite{circles}, in collaboration with automobile manufacturers, tested AI and machine learning in actual traffic. Their preliminary simulations indicated that AI-equipped vehicles could mitigate phantom jams and save energy at least 10\% in real-world traffic applications. Although RL-based algorithms appear promising, they often suffer from overfitting to training scenarios and the sim-to-real gap, consequently making it challenging to generalize to real-world traffic. The relative ease of generalizability in our approach is due to its objective-based nature, as opposed to the predominantly policy-based focus found in much of car-following literature.

In this work we concentrate on the situation where the traffic oscillations are exclusively induced by human driving behaviors. Therefore, we adopt the setting in the Tadaki experiment, where a varying number of vehicles traverse a circular single-lane road of a given circumference.  Our work differs from previous works mostly from a methodological perspective.  First,  we adopt a realistic human driving behavior model \cite{ComputationalFramework}, in which driving policy is derived by utility maximization individually. The model parameters are calibrated and fixed using vehicle trajectory data from the Sugiyama experiment \cite{CalibrationPaper}. Traffic dynamics are therefore completely endogenized in our approach without tuning any free parameters.  Second, by taking advantage of emerging technology, such as connectivity and edge computing, we continue the trend of deploying CAVs to improve traffic flow and smoothness.  However, we treat the control of the CAV by the traffic control authority as a computational mechanism design problem, in which all traffic is naturally modeled as a simulated game among all human-driven vehicles and system-controlled CAVs. Our behavior model suggests a very simple one-parameter family of control law that can be systematically optimized.  Tradeoff between boosting traffic flow and suppressing traffic oscillation can be conveniently handled by optimizing a properly defined system-level objective function, resulting in the desired efficient frontier. Numerically,  our results demonstrate that both traffic efficiency and smoothness can be improved substantially.

The simplicity of the control law deployed in our approach also affords us an opportunity to elucidate its precise mathematical nature.  In a companion work \cite{Bifurcation} a systematic bifurcation analysis is performed for the specific case when all vehicles are CAVs (indirectly enforceable via variable speed advisory) and all heterogeneity and noise are switched off.  It is shown there that the optimal control law in such a ``clean'' setting coincides with the 2-dimensional bifurcation curve of the system in the $(\rho,v^*)$-plane, where $\rho$ is the vehicle density and $v^*$ the recommended speed. This implies that the control law found via computational mechanism design, though distorted by driver heterogeneity and state evolution noise, has its mathematical origins from bifurcation theory.

The rest of the paper is organized as follows. In the next section we briefly review the human driving behavior modeling and calibration recently introduced in \cite{ComputationalFramework} and \cite{CalibrationPaper}. Even though the model calibration is done using traffic trajectory data from the Sugiyama experiment, we demonstrate that the same model can be generalized into the new settings of the Tadaki experiment. This verifies that the density dependence and fundamental diagram are well reproduced in our simulation. In Section \ref{Mechanism_design} we formulate the traffic control problem at the system level as a computational mechanism design and specify a very simple family of control policies. We then derive the efficient frontier of the optimal control and quantify the degree to which the controlled traffic efficiency and smoothness can be improved in the next two sections, along with intuitions for understanding the derived results. In Section \ref{Optimal_control_policy}, the case involving one CAV is examined and presented with some technical details. In Section \ref{Multi-CAV}, we present results for cases of multiple CAVs. We envision three possibilities for implementing the traffic control system in reality in Section \ref{System_realization}. We summarize and conclude in the last section.

\section{\bf The Human Driving Behavior Model and Calibration}\label{BehavioralModeling}
In \cite{ComputationalFramework} we proposed a heuristics-based driving decision algorithm called {\it adaptiveSeek}. In this approach a Markov game with continuous actions and simultaneous moves was used as the modeling framework. The solution concept {\it adaptiveSeek} was simplified from a rigorous game theory-based approach dubbed {\it betaNash} by relaxing some of the theoretical requirements that were too restrictive, both conceptually and computationally. It was shown explicitly that {\it adaptiveSeek} can approximate the sub-game perfect Nash equilibrium solutions well while taking into account human drivers' bounded rationality, as illustrated in a situation of mandatory lane changes on a double-lane highway.  It was also shown in \cite{CalibrationPaper} that the model parameters for traffic on a circular road in the Sugiyama setting can be quantitatively calibrated using observed vehicle trajectory data.  Furthermore, it will be explicitly demonstrated that the calibrated model generalizes nicely to the setting of the Tadaki experiment \cite{Tadaki13}, where vehicle density is varied, in the sense that it can reproduce nearly all the observed collective phenomena well (see Subsection E). Therefore, we will use {\it adaptiveSeek} as the decision-making model for human driving behaviors. 

\subsection{The Behavioral Model Specification}
Here we recapitulate the decision-making algorithm {\it adaptiveSeek} introduced in \cite{ComputationalFramework,CalibrationPaper}. Specific considerations and justifications can be found in the original papers. Only longitudinal dynamics are explicitly modeled.
\subsubsection{The traffic setting}
There are $N$ vehicles in the system traversing on a circular road of circumference $C$. The vehicle ordering is chosen so that vehicle $i$ is always behind vehicle $i+1$. Periodic boundary conditions are imposed, i.e. all position variables $x_{i,t}$ (and headway) should be understood as $\pmod{C}$, and the $(N+1)$-th vehicle is identified as the first vehicle.  For simplicity, we use {\it agent} to refer either vehicle, driver,  or a combination of both.
\subsubsection{Driving decision-making}
There are two kinds of state variables that are conceptually distinct. The first kind is for characterizing the kinematics of vehicles, and the second for drivers' decision-making. The kinematic state for vehicle $i$ is defined by a vector $\xi_{i,t}=(x_{i,t},v_{i,t},a_{i,t})^\top$,  whose components represent the position, velocity and acceleration of the vehicle at time $t$, respectively.  The decision-making state by the driver of vehicle $i$ at time $t$ is defined by $s_{i,t}=(\xi_{i,t}^\top,\xi_{j,t}^\top)^\top$\footnote{Conceptually, $\xi$'s components appearing in $s_{i,t}$ could be different from the $\xi$'s for characterizing vehicle's motion, due to driver's estimation error. We will ignore such differences in this work for simplicity.}, involving the ego vehicle $i$ and the leading vehicle $j\equiv i+1$. 

The kinematic state evolution can be written as
\begin{equation}
	\begin{cases}
		x_{i,t+1}&= x_{i,t}+v_{i,t}\,\Delta t +\mu^x_{i,t+1}\,\,\,\pmod{C}\, ,\\
		v_{i,t+1}&=v_{i,t}+a_{i,t}\,\Delta t+\mu^v_{i,t+1}\, , \\
		a_{i,t+1} &=\gamma\,a_{i,t}+ \big(u_{i,t}- \gamma\,u_{i,t-1} \big)+\mu^a_{i,t+1},
	\end{cases}
	\label{VehicleDynamics}
\end{equation}
with $u_{i,t}$ being the control input (or action) of the driver. The noises are all IID: $\mu^x_{i,t+1}\sim N(0,\sigma_x^2)$, $\mu^v_{i,t+1}\sim N(0,\sigma_v^2)$,  and $\mu^a_{i,t+1}\sim N(0,\sigma_a^2)$. Note that the acceleration is modeled as an AR(1) process in order to capture its stickiness in vehicle dynamics.

Decision-making is driven by utility maximization. In a game setting, the utility is not only a function of its own control input $u_{i,t}$ but is also dependent on control inputs from all other interacting neighboring agents $u_{-i,t}$. In order to more closely mimic the bounded rationality of human driving, we must relax the strong assumptions associated with dynamic Nash equilibria. Therefore, additional behavioral assumptions are made, which are appropriate for car-following: constant action for the ego agent ($u_{i,t}=u$); zero action for other agents ($u_{-i,t}=0$). Thus, the driving policy for driver $i$ can be expressed as
\begin{equation}
	u^*_{i,t}(s_{i,t})=\argmax_{u\in[u_\text{min},u_\text{max}]} 
	U^\text{eff}_{i,t}(u|s_{i,t})\, . \label{BestResponseTilde}
\end{equation}
The effective utility typically consists of several utility components, each of which takes care of a specific aspect of driving preference. It is
further related by multiple per-period utilities evaluated at anticipated future states within the planning horizon $H+1$: $\forall h\in\{0,\cdots,H\}$
\begin{equation}
	U^\text{eff}_{i,t}(u|s_{i,t})=\sum_{k}\, w_{i,k} \,g_k\Big[\cup_{h=0}^H U_{i,t}^{(k)}(u|\hat{s}_{i,h})\Big]_{\hat{s}_{i,0}=s_{i,t}}\, ,
	\label{EffectiveUtility}
\end{equation}
where $U_{i,t}^{(k)}$ is the $k$-th component of the per-period utility with weight $w_{i,k}$. Instead of the standard cumulative utility form, we use transformed utility with $g_k[...]$. This transformation is necessary for better fitting human driving trajectories. Starting from the current state $\hat{s}_{i,0}=s_{i,t}$, the anticipated future state for the ego agent $i$ evolves according to
\begin{equation}
	\begin{cases}
		\hat{x}_{i,h+1}&=\hat{x}_{i,h}+\hat{v}_{i,h}\,\Delta t \,\,\,\pmod{C}\\
		\hat{v}_{i,h+1}&=\hat{v}_{i,h}+\hat{a}_{i,h}\,\Delta t\\
		\hat{a}_{i,h+1}&=u
	\end{cases}
	\label{anticipation_evolution}
\end{equation}
and similarly for other interacting agents ($-i$) with zero action.

Conceptually, it is important to distinguish the two state evolutions defined in Eq.(\ref{VehicleDynamics}) and Eq.(\ref{anticipation_evolution}). The former models the mechanical realization for vehicle $i$ given driver $i$'s control inputs, whereas the latter models driver $i$'s mental anticipation of future states for all relevant agents. 

\subsection{Utility function and its parameterization}
For our car-following setting we only need three utility components. In the following, we utilize the available information to the fullest, in the sense that by knowing $\hat{s}_{i,h}=(\hat{x}_{i,h},\hat{v}_{i,h},\hat{a}_{i,h})$, we also know $\hat{x}_{i,h+1}=\hat{x}_{i,h}+\hat{v}_{i,h}\Delta t$ and $\hat{v}_{i,h+1}=\hat{v}_{i,h}+\hat{a}_{i,h}\Delta t$.
\subsubsection*{Moving forward reward at a desired speed}\label{Moving_forward}
The first component represents the intention to move forward along the circular road at the desired speed:
\begin{equation}
	U_{i, t}^{(1)}(u|\hat{s}_{i,h})=\exp\Big(-\Big(\frac{\hat{v}_{i, h+1}+u\,\Delta t- v_i^*}{\kappa_i^{(1)} v_i^*}\Big)^{2}\Big)\, ,
\end{equation}
where $v_i^*$ is the ideal speed and $\kappa_i^{(1)}$ controls the degree by which agent $i$ likes to be close to its ideal speed. 

The anticipation related transformation function for this first component is given by 
$$g_1\big[\cup_{h=0}^H\,U_{i, t}^{(1)}(u\big|\hat{s}_{i,h})\big]=U_{i, t}^{(1)}(u|\hat{s}_{i,h})\big|_{h=0} \, ,$$
{\it i.e.} for all $H+1$ items, only the first in the sequence matters.

\subsubsection*{Moving backward penalty}\label{Moving_backward}
The second component is a penalty for moving backward:
\begin{equation}
 U_{i, t}^{(2)}(u|\hat{s}_{i,h})=\exp\Big(-\kappa_v^{(2)}\big(\hat{v}_{i, h+1}+u\,\Delta t +\kappa_0^{(2)}\big)\Big)\, .
\end{equation}
This term is needed in order to prevent vehicle velocity from being persistently negative. Also, $g_2$ is chosen similarly as $g_1$:
$$g_2\big[\cup_{h=0}^H\,U_{i, t}^{(2)}(u\big|\hat{s}_{i,h})\big]=U_{i, t}^{(2)}(u|\hat{s}_{i,h})\big|_{h=0} \, .$$

\subsubsection*{Pairwise collision penalty}\label{Pairwise_collision}
The third component represents the subjective risk perceived by the driver for one-on-one collision with another vehicle.  Let $L_i$ denote the length of vehicle $i$, and let $\mathcal{F}(x)=\exp{( -x^2-2x )}$. For vehicle $i$'s front collision, we choose
\begin{equation}
	U_{i, t}^{(3)}(u|\hat{s}_{i,h})=\left\{
	\begin{array}{ll}
		1, & \Delta x_{i, j, h} \leq 0 \\ [6pt]
		\mathcal{F} \Big(\displaystyle\frac{\Delta x_{i, j, h}}{\delta_{i,j, h}}\Big), & 0<\Delta x_{i, j, h}
	\end{array}\right. 
	\label{PairwiseCollisionPenalty}
\end{equation}
where $\Delta x_{i,j,h}=(\hat{x}_{j,h+1}+\hat{v}_{j,h+1}\Delta t-L_{j}/2)-(\hat{x}_{i,h+1}+\hat{v}_{i,h+1}\Delta t+L_i/2)$ is the bumper-to-bumper distance of vehicle $i$ to its leading vehicle $j$. It is natural to choose the front scale parameter to be speed dependent: $\delta_{i,j,h} = \kappa_{i,c}^{(3)}+\kappa_{i,v}^{(3)}|\hat{v}_{i,h+1}+u\Delta t | + \kappa_{i,d}^{(3)}\max\{\hat{v}_{i,h+1}+u\Delta t-\hat{v}_{j,h+1},0\}$. The $\kappa_{i,d}^{(3)}$ term is needed for braking when the vehicle ahead is slower.  Because of the discouragement for moving backward in $U_{i, t}^{(2)}$ we do not need to explicitly consider the rear collision penalty here.
The anticipation-related transform function for the third component is given by 
$$g_3\big[\cup_{h=0}^H\,U_{i, t}^{(3)}(u\big|\hat{s}_{i,h})\big]=\max_{h\in\{0,\cdots,H\}} U_{i,h}^{(3)}\big(u|\hat{s}_{i,h}\big)\, .$$

\subsection{Model Parameters}
Our goal is to have a realistic driving behavior model for the Tadaki setting.  Ideally, we should have calibrated all the model parameters if we had vehicle trajectory data from the Tadaki experiment.  Since this is not the case, the closest data we can find is that of the Sugiyama experiment. Therefore, we adopted the same behavior model and its model parameters from the Sugiyama setting \cite{CalibrationPaper}.  We then average over all 22 Sugiyama agents to arrive the following common model parameters for this study\footnote{In order to avoid occasional collisions for very long simulation sessions, the risk premium parameters are slightly enlarged,  while acceleration noise is reduced.  }: $v^*=10.49$ (m/s),  $\kappa^{(1)}=0.7$,  $w^{(1)} =1$; $\kappa_v^{(2)}=10$,  $\kappa_0^{(2)}=0.25$ (m/s),  $w^{(2)}=-1$; $\kappa_c^{(3)}=0.6$ (m),  $\kappa_v^{(3)}=0.3$ (s),  $\kappa_d^{(3)}=1.0$ (s), and $w^{(3)}=-10$.  We further choose $\Delta t=1/3$ (s), $\gamma=0.7$, and $H=3$ in our specific context, which leads to a look-ahead time horizon of $(H+1)\Delta t=4/3$ (s).  All vehicles have the same length: $L=3.9$ (m).  The standard deviations of state evolution noise are $\sigma_x=0.05$ (m), $\sigma_v=0.1$ (m/s), and $\sigma_a=0.1$ (m/s$^2$). A small amount of heterogeneity at $5\%$ level is then added to individual agent's preference (only on the most sensitive ones: $v^*$, $\kappa_v$ and $\sigma_a$) relative to the averages. Control input is limited as $u\in\mathcal{U}=[u_\text{min},u_\text{max}]=[-6,4]$ (m/s$^2$). 

Once these parameters are chosen, our driving behavior model is completely fixed for the Tadaki setting. All driving behaviors are endogenized, purely induced by the driver's decision-making states and specific realization of the noise terms.
\subsection{Traffic Dynamics}\label{TrafficDynamics}
For mathematical convenience we approximate \eqref{BestResponseTilde} via the following Boltzmann weighted average:
\begin{equation}
	\bar{u}_{i,t}(s_t)\equiv \sum_{u} u P_{i,t}(u|s_{i,t})\, ,
	\label{u_bar}
\end{equation}
with $P_{i,t}(u|s_{i,t})\propto\exp[\lambda\, U^\text{eff}_{i,t}(u|s_{i,t};h)]$.  When $\lambda$ is positive and large, $\bar{u}_{i,t}(s_t)\rightarrow u^*_{i,t}(s_{i,t})$.  Thus, we can regard the control input in Eq.(\ref{u_bar}) as a regularized version of Eq.(\ref{BestResponseTilde}),  such that $\bar{u}$ has a continuous support and is differentiable with respect to model parameters, even when the utility maximization is done via a grid search.\footnote{The grid we choose has 41 points for $u\in[-6,4]$ with $\lambda=200$.}

The endogenized state evolution in Eq.\eqref{VehicleDynamics} for vehicle $i$ can then be expressed as follows: $\forall i\in\{1,\cdots,N\}$
\begin{equation}
	\begin{cases}
		x_{i,t+1}&= x_{i,t}+v_{i,t}\,\Delta t +\mu^x_{i,t+1}\,\,\,\pmod{C}\, ,\\
		v_{i,t+1}&=v_{i,t}+a_{i,t}\,\Delta t+\mu^v_{i,t+1}\, , \\
		a_{i,t+1} &=\gamma\,a_{i,t}+ \big(\bar{u}_{i,t}(s_{i,t})- \gamma\,\bar{u}_{i,t-1}(s_{i,t-1}) \big)+\mu^a_{i,t+1},
	\end{cases}
	\label{TrafficDynamics}
\end{equation}
The above state evolution is executed by all agents in parallel step-by-step, as implied by the Markov game setting. In so doing, every agent is treated symmetrically, in the sense that each vehicle is simultaneously an ego vehicle in its own state evolution, but also can appear as a surrounding vehicle in other agents' utility calculation. Collectively, Eq.(\ref{TrafficDynamics}) defines traffic dynamics.

\subsection{Generalization to the Tadaki Experiment}\label{Tadaki_Verification}
We now demonstrate that our behavioral model, which we calibrated using the Sugiyama experiment data, can reproduce the observations in the setting of Tadaki et al \cite{Tadaki13} well. Differing from the  original Sugiyama setting, the Tadaki experiment utilized a bigger circular ring in Nagoya Dome (a baseball stadium) with a 314 (m) circumference. To study density dependence, the number of cars were varied from 10 to 40. Vehicles were all homogenized as Toyota Vitz. Drivers were all college students.

Generally, the timing of the spontaneous formation of the stop-and-go wave is random and our simulations are inevitably finite. Therefore, it is not possible for us to precisely create the same conditions in the simulation as those in the Tadaki experiment. In addition, it is impossible to know the amount of heterogeneity and noise level involved in the original experiment, so there will be differences there too. Unfortunately,  some of the simulation results can depend on these details. To avoid ambiguity,  we adhere to the following procedure throughout the rest of the paper. Driver heterogeneity is randomly generated (through randomized preference parameters) and then fixed. Each simulation run consists of 3000 evolution steps, i.e. 1000 seconds of traffic. It starts with all vehicles evenly spaced and at rest. At $t=10$ (s), one of the vehicles is "kicked", i.e. it brakes at $-1.0$ m/s$^2$ for the next 6 seconds, as long as the velocity is positive. This seeds the possibility of stop-and-go for each run. If $N$ is sufficiently small, this initial disturbance will fizzle out and the system stays in free-flow phase.  If $N$ is large enough, the disturbance will grow to a sustainable stop-and-go wave, putting the system in the jammed-flow phase.  The measurement is done within the window $t\in[200,1000]$ (s). Finally, each run is repeated multiple times, with different random seeds for the state evolution noise.

\begin{figure}[!h]
\centering
\includegraphics[width=0.44\textwidth]{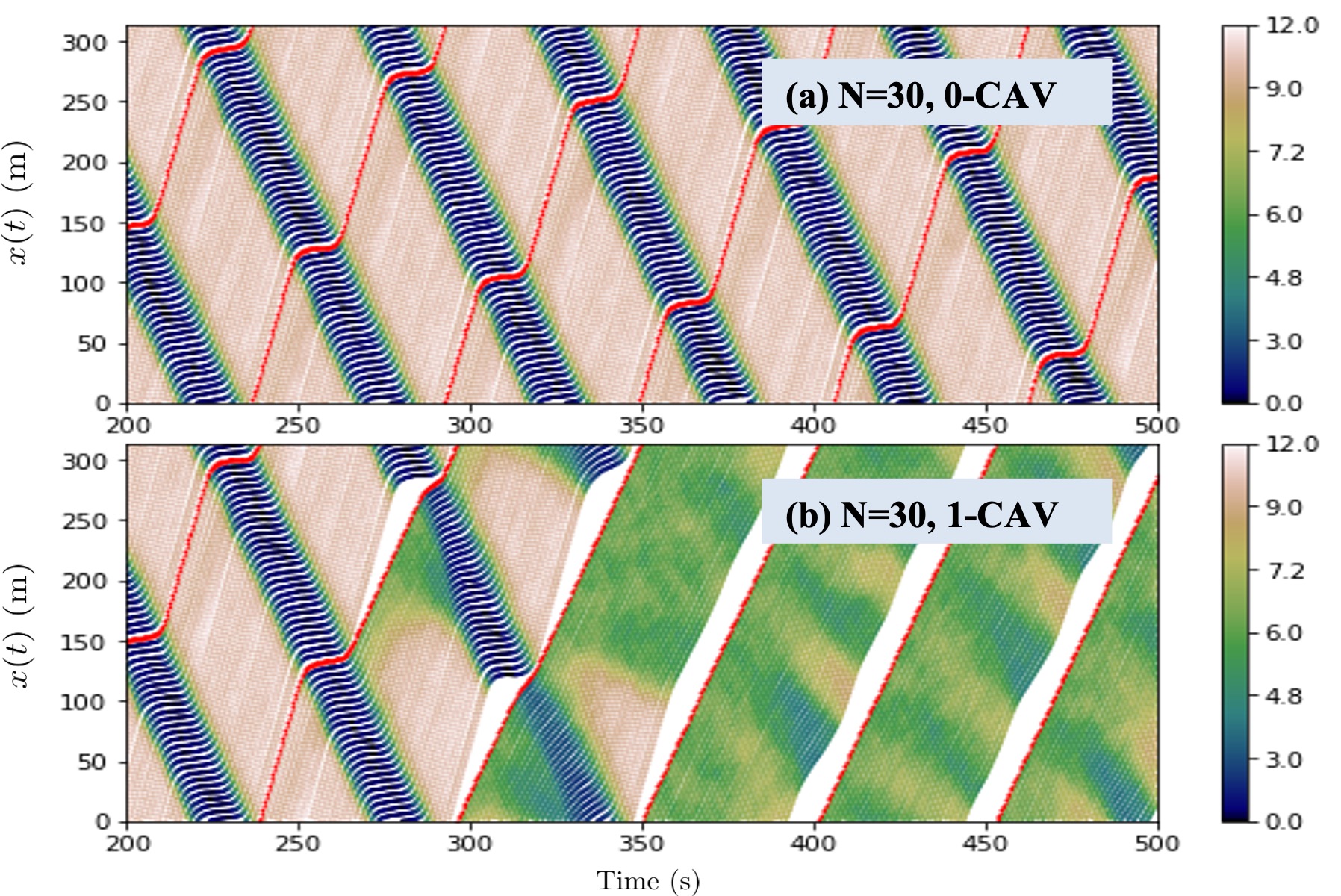}
\caption{(a) When all vehicles ($N=30$) are human driven: stop-and-go wave is clearly visible; (b) When vehicle $\#1$ is a system-controlled slow CAV: stop-and-go wave is dissipated once the CAV (with $v_1^*=6.1$ m/s) is switched on at $t=250$ (s).  The red line depicts the trajectory of vehicle \#1.  The color legend represents instantaneous speed of the vehicle in m/s. }
\label{trajectory_plots_N30_0or1CAV}
\end{figure}

To show that our driving behavior model is reasonable we first verify two qualitative results: 1) spontaneous formation of stop-and-go waves in our simulation at roughly the right vehicle density \cite{Tadaki13}; and 2) the waves can be dissipated with a system-controlled CAV \cite{Stern2018} whose cruising speed is set at a lower value than a typical driver's ideal speed.  In Fig.\ref{trajectory_plots_N30_0or1CAV} we display simulated vehicle trajectories at $N=30$ with and without the system controlled CAV.  When all vehicles are human driven stop-and-go waves are generated and sustained (see Fig.\ref{trajectory_plots_N30_0or1CAV}(a)).  Quantitative similarity between our Fig.\ref{trajectory_plots_N30_0or1CAV}(a) and Fig.4(a) in \cite{Tadaki13} demonstrates that {\it adaptiveSeek} is able to closely model the observed human driving behaviors in the Tadaki setting.  Important characteristics, such as critical vehicle density, average speed,  speed range, queue length of the slow moving cluster, and backward traveling slope of the shockwave,  are all closely reproduced.  Once vehicle \#1 turns into a CAV at $t=250$ (s) and its ideal speed is kept sufficiently slow, the wave is gradually dissipated (see Fig.\ref{trajectory_plots_N30_0or1CAV}(b)), similar to Fig.3 and Fig.4 in \cite{Stern2018}.  The timing of the dissipation process is generally random and appears to be quicker statistically when CAV's ideal speed is lower.

\begin{figure}[!h]
	\centering
		\includegraphics[width=0.44\textwidth ]{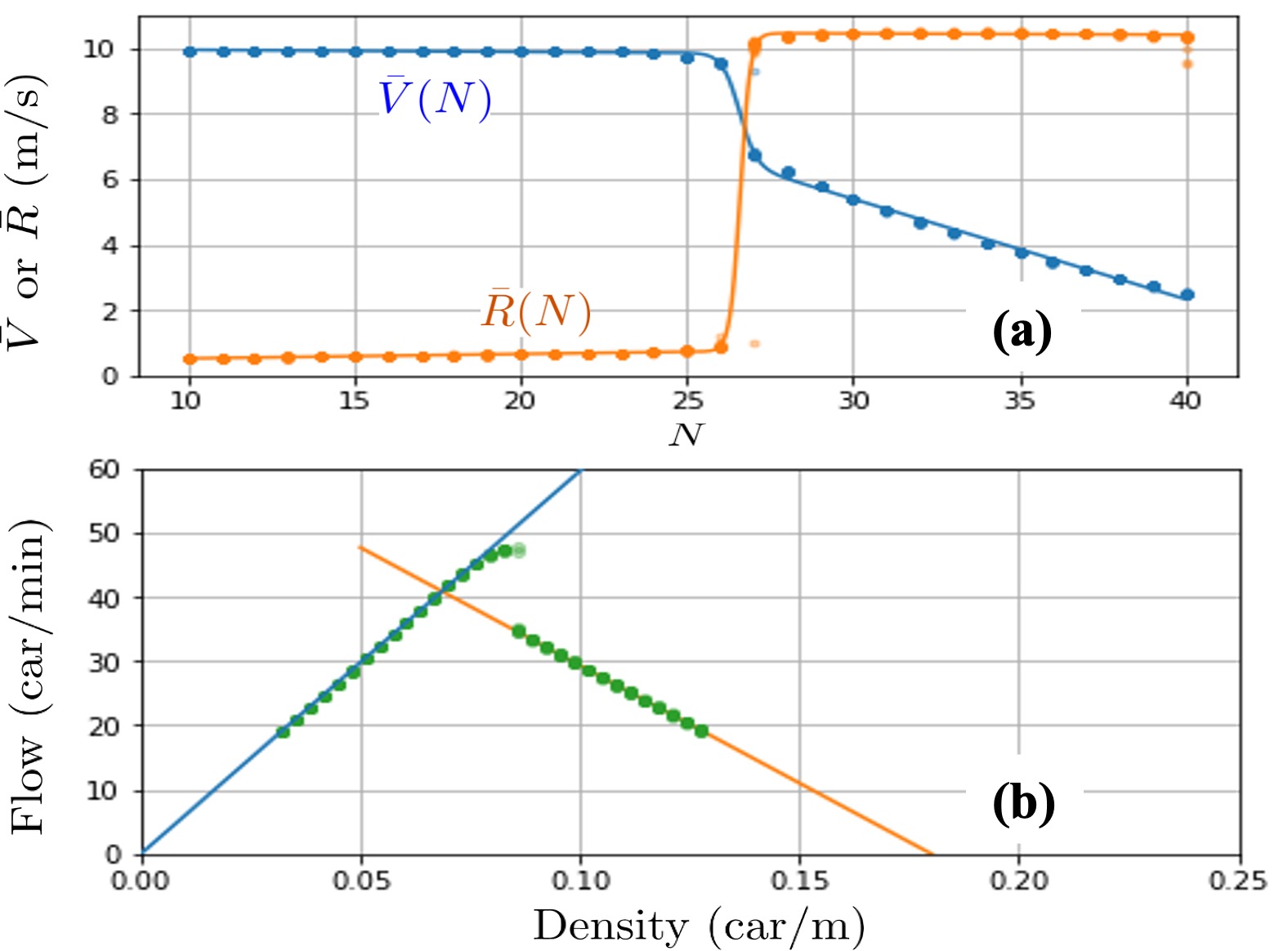}
	\caption{When all vehicles are human driven: (a) Long-run mean of average-speed (blue) and speed-range (orange) as functions of the number of vehicles traversing the ring road, with smooth curves fitted to guide the eye; (b) The corresponding fundamental diagram, with straight lines fitted to data below (blue) and above (orange) the phase transition density $\sim$ 0.086 (car/m), or $N_c=27$, with the slope of the blue line close to the average ideal speed.}
	\label{ReproducedTadaki}
\end{figure}

We then verify two quantitative results observed in the Tadaki experiment in Figure 5 and Figure 6 of their paper respectively: 1) long-run mean of average-speed vs the number of cars; 2) fundamental diagram, i.e. flow versus density.  In Fig.\ref{ReproducedTadaki}(a) we show both the simulated long-run mean of average-speed and speed-range as functions of the number of vehicles on the circular road. In Fig.\ref{ReproducedTadaki}(b) we show the simulated fundamental diagram.  The system is able to sustain stop-and-go waves starting at $N=27$. \footnote{If the kick were not applied, the transition point would have shifted upwards to $N=28$, with otherwise the same simulation runs.} Apart from the fact that the phase transition from free flow to jammed flow is sharper,  almost all quantitative results are well reproduced in our simulation. It is also clear that speed-range would serve as a better order parameter for signaling the phase transition mentioned in \cite{Tadaki13}, because it is sharply different in the free-flow phase versus the jammed-flow phase. 

\section{\bf Traffic Control as a Mechanism Design Problem}\label{Mechanism_design}
For composition simplicity we concentrate explicitly on the case of one CAV.  Generalization to cases with multiple CAVs is straightforward. In the remainder of this paper all vehicles will be modeled by {\it adaptiveSeek}. However, we distinguish the system controlled CAV, the first vehicle ($i=1$), from all others vehicles ($i\in\{2, ..., N\}$) that are human driven. This means we will treat almost all utility parameters as those of the calibrated vehicles, fixed with the proper heterogeneity at the individual level. The one exception will be the ideal speed for the CAV, $v_1^*$, which will be treated as a system-controlled function. We will describe this in more detail later.

As suggested in \cite{ComputationalFramework}, mechanism design can be used for prototyping and optimizing better/smarter traffic systems. The idea is conceptually straightforward: the principal (or traffic authority) lets all the relevant agents (or drivers) play a driving game under a set of traffic rules/settings and adjust the rules/settings so that a chosen aggregate objective is maximized. Game-playing would allow agents to incorporate their strategic reactions to changes in the traffic rules or setting, as well as to other agents at an appropriate level. Therefore, incentive compatibility will be properly maintained via individual utility maximizations, which is key for mechanism design \cite{Fudenburg96}. The process of design and optimization is called computational because it often involves simulating the game repeatedly, where the traffic rules/settings are varied. In our specific case, we do not change traffic rules; we only vary the setting for human-driven vehicles. The setting variation is indirectly mediated by the presence of the CAV whose behavior is controlled by the traffic control authority.

We have our dual-objective in mind when designing the traffic control system: traffic throughput and smoothness. These objectives are the bare minimum for this system\footnote{Other objectives, such as those related to safety, improving fuel consumption, and reducing various emissions, can be also added, provided that the relevant measurements are available. For example, safety can be quantified by the aggregate collision penalty in our approach.}, as emphasizing one objective while ignoring the other could potentially lead to nonsensical conclusion. For example, holding all vehicles still will certainly smooth out any waves,  but also defeat the purpose of traffic.

The first step in mechanism design is to find the appropriate measures for these objectives. For a good measure of throughput we choose the average-speed of all vehicles
\begin{equation}
	V(t)=\frac{1}{N}\sum_{i=1}^N v_{i,t}\, .
\end{equation}
To quantify the traffic variability we choose the speed-range among all vehicles, as it is the most sensitive measure possible:\footnote{The speed variance among all vehicles would almost be as good.}
\begin{equation}
	R(t)=\max_{i=1}^N v_{i,t}- \min_{i=1}^N v_{i,t}\, .
\end{equation}
We use capital letters here to emphasize the fact that the above two variables are defined in aggregate at the system level.

The second step is to define an objective for the control system, whose optimization provides the optimal control policy. Since the two objectives can be potentially conflicting, in the sense that optimizing one objective may not imply optimizing the other, we introduce a tradeoff parameter $\omega>0$ that weighs the relative importance of the throughput (higher is better) and speed variation (lower is better)
\begin{equation}
	\Pi[v_1^*(S(t)|\kappa)|N,\omega]=\frac{1}{T}\sum_{t=1}^T\Big(V(t)-\omega R(t)\Big)\, ,
\end{equation}
where $T$ is a long time horizon. In the above definition the system control is implemented through the ideal speed for the CAV ($i=1$), $v_1^*(S(t)|\kappa)$, with $S(t)$ being a state variable defined at system level that captures some aspect of the collective motion of the traffic, and $\kappa$ the parameter that characterizes the control policy.  How elaborate we should make $S(t)$ requires careful deliberation.  This is because we need to trade off theoretical control efficiency with practical system implementation.  Intuitively, a reasonable compromise is to choose the system-level state as an appropriately chosen local\footnote{The locality here should not be understood at a micro level or the immediate neighborhood of the CAV. Rather, it should be defined at a length-scale appropriate for the natural formation of stop-and-go waves in the specific situation. In our current context, it is as long as the circumference of the circular road. Generally, the precise definition should be explored carefully as a part of the optimization process by the traffic control system.} measure of vehicle density averaged over a number of periods:
\begin{equation}
	S(t)=\frac{1}{\tau}\sum_{t'=1}^{\tau} \rho(t-t')\, .
\end{equation}
Because of the specific setting of the Tadaki experiment, the vehicle density is really a constant for each given $N$.  Consequently,  the system-level control function is also a constant:
\begin{equation}
	v_1^*(S(t)|\kappa))=\kappa\, .
	\label{Cruising_speed_controller}
\end{equation}
This simplest possible functional form turns out to be sufficient for our purpose. Furthermore, this simplicity will greatly facilitate implementation, which we will discuss in Section \ref{System_realization}.  

The third step is to specify the multi-agent dynamic evolution of the system. The longitudinal evolutions of position and speed, and dynamics for vehicle $i\in\{1, 2, ..., N\}$ are described by Eq.(\ref{TrafficDynamics}), with all the individual level control input governed by Eq.{\ref{u_bar}).  For the system controlled CAV the optimal control input is governed by
\begin{equation}
	\bar{u}_{1,t}^*\big(s_{1,t}|\tilde{\theta}_1,\kappa\big)=
	\text{softmax}_{u\in\mathcal{U}}U_{1,t}\big(u|s_{1,t};\tilde{\theta}_1, v_1^*=\kappa\big)\, ,
	\label{CAV_controller}
\end{equation}
where $\tilde{\theta}_1$ represents all the other components of $\theta_1$ (full set of preference parameters for agent $1$) except the ideal speed.  It is also interesting to point out that our CAV controller defined in Eq.(\ref{Cruising_speed_controller}, \ref{CAV_controller}) is essentially a more parsimonious version of the high-level controller introduced in \cite{Stern2018} called the FollowerStopper controller. Since our driver behavioral model is calibrated using the trajectory data from the Sugiyama experiment, there are no free parameters in our CAV controller other than the traffic control variable $v_1^*=\kappa$. 

As the last step to complete the formulation, we achieve our design objective with the following optimization:
\begin{equation}
	\kappa^*(N,\omega)=\text{argmax}_\kappa\Pi[\kappa|N,\omega]\, .
\end{equation}
Because the long-run mean for average-speed and speed-range no longer depends on $\omega$ with this particular control policy, the system-level objective function becomes simplified: $\Pi(\kappa|N,\omega)=\bar{V}(\kappa|N)-\omega\bar{R}(\kappa|N)$. This simplification makes the efficient frontier calculation much easier. In more complicated cases,  the control function might be parameterized as a specific form of neural networks whose training can be achieved by applying reinforcement learning approach, such as in \cite{Freidieh2018}.

\section{\bf Optimal Control Policy and Efficient Frontier}\label{Optimal_control_policy}
We now derive a system-level optimal control policy, $\kappa^*(N,\omega)$ defined in Eq.({\ref{CAV_controller}). Specifically, we use $N=30$ with one system-controlled CAV as the illustrative example. The task is to search for the optimal value of the ideal speed for the CAV so that the objective is maximized. To this end we scan a range of values for $v_1^*=\kappa$ with an increment of $\Delta\kappa=0.01$ m/s. At each $\kappa$ value, one hundred simulations are run with different random seeds. We need a large number of runs because there exists a transition region where variations for both average-speed and speed-range are high. 
We continue to follow the simulation procedure outlined in subsection \ref{Tadaki_Verification}.
The CAV control is always switched on at $t=50$ (s) in every run.  The simulation results are displayed in Fig.\ref{kappa_scan_N30}. We can recognize three distinct phases, each corresponding to a different combination of average-speed and speed-range.  These phases are separated by two values on the $\kappa$-axis, roughly located at $\kappa_\text{\tiny{L}}\sim 2.5$ (m/s) and $\kappa_\text{\tiny{H}}\sim 6.3$ (m/s). When $\kappa>\kappa_\text{\tiny{H}}$ the traffic is in jammed flow (low average-speed and high speed-range). When $\kappa_\text{\tiny{L}}<\kappa<\kappa_\text{\tiny{H}}$ the traffic is in free flow (high average-speed and low speed-range). When $\kappa<\kappa_\text{\tiny{L}}$ the traffic is in suppressed flow (both average-speed and speed-range are limited by $\kappa$).  Note that only the transition from jammed-flow to free-flow is sharp, whereas the transition from free-flow to suppressed-flow is gradual.

\begin{figure}[!h]
	\centering
	\includegraphics[width=0.46\textwidth]{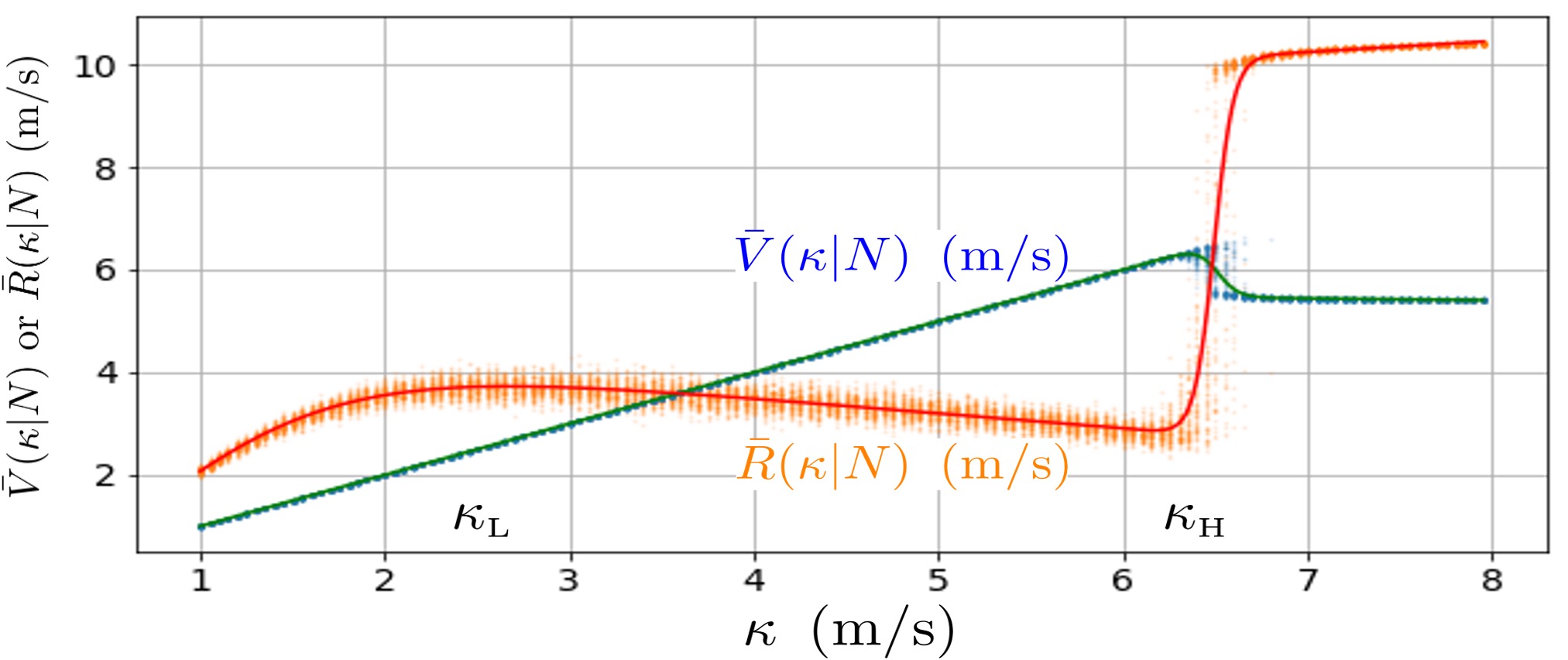}
	\caption{Simulation results with one CAV at $N=30$ and a range of $\kappa$ values: Long-run mean of average-speed and speed-range as functions of the ideal speed of the CAV. Three distinct phases of traffic flow, roughly delineated by $\kappa_\text{\tiny{L}}\sim 2.5$ (m/s) and $\kappa_\text{\tiny{H}}\sim 6.3$ (m/s), can be identified: suppressed-flow ($\kappa<\kappa_\text{\tiny{L}}$), free-flow ($\kappa_\text{\tiny{L}}<\kappa<\kappa_\text{\tiny{H}}$), and jammed-flow($\kappa>\kappa_\text{\tiny{H}}$).}
	\label{kappa_scan_N30}
\end{figure}
Qualitatively, the shape of Fig.\ref{kappa_scan_N30} can be understood as follows. When it has a high ideal speed ($\kappa>\kappa_\text{\tiny{H}}$), the CAV is behaving like a usual human driver. Therefore, the entire system can suffer from the stop-and-go waves as if all vehicles involved were human driven. When the ideal speed of the CAV is lowered to an intermediate range ($\kappa_\text{\tiny{L}}<\kappa<\kappa_\text{\tiny{H}}$), the CAV breaks away from the chasing behavior that a typical human driver would possess. Consequently, the wave will lose its rhythm so that it can no longer sustain itself in this phase.  On the other hand, since $\kappa$ sets the overall traveling pace of the fleet, it is natural to expect that the average-speed of the traffic is proportional to $\kappa$ when $\kappa<\kappa_\text{\tiny{H}}$. A curious simulation outcome is that the speed-range becomes higher when $\kappa$ is lowered in the free-flow phase of Fig.\ref{kappa_scan_N30}. This phenomenon is related to the fact that human-driven vehicles are having a harder time keeping pace with the CAV at a slower constant speed without being ``tripped up'' occasionally (see Fig.\ref{trajectory_N30_slower}). Consistent with our simulation results, a similar ``trip-up'' phenomenon was also observed in Fig.3 and Fig.4 (Experiment A and B, respectively) in \cite{Stern2018} with actual human drivers and real cars. Lastly, when the ideal speed of the CAV is further lowered to very small values ($\kappa<\kappa_\text{\tiny{L}}$), there is little room left for human-driven vehicles to even trip up. This is where the suppressed-flow phase sets in. In this phase both average-speed and speed-range are suppressed by the ideal speed of the CAV. 

\begin{figure}[!h]
	\centering
		\includegraphics[width=0.44\textwidth]{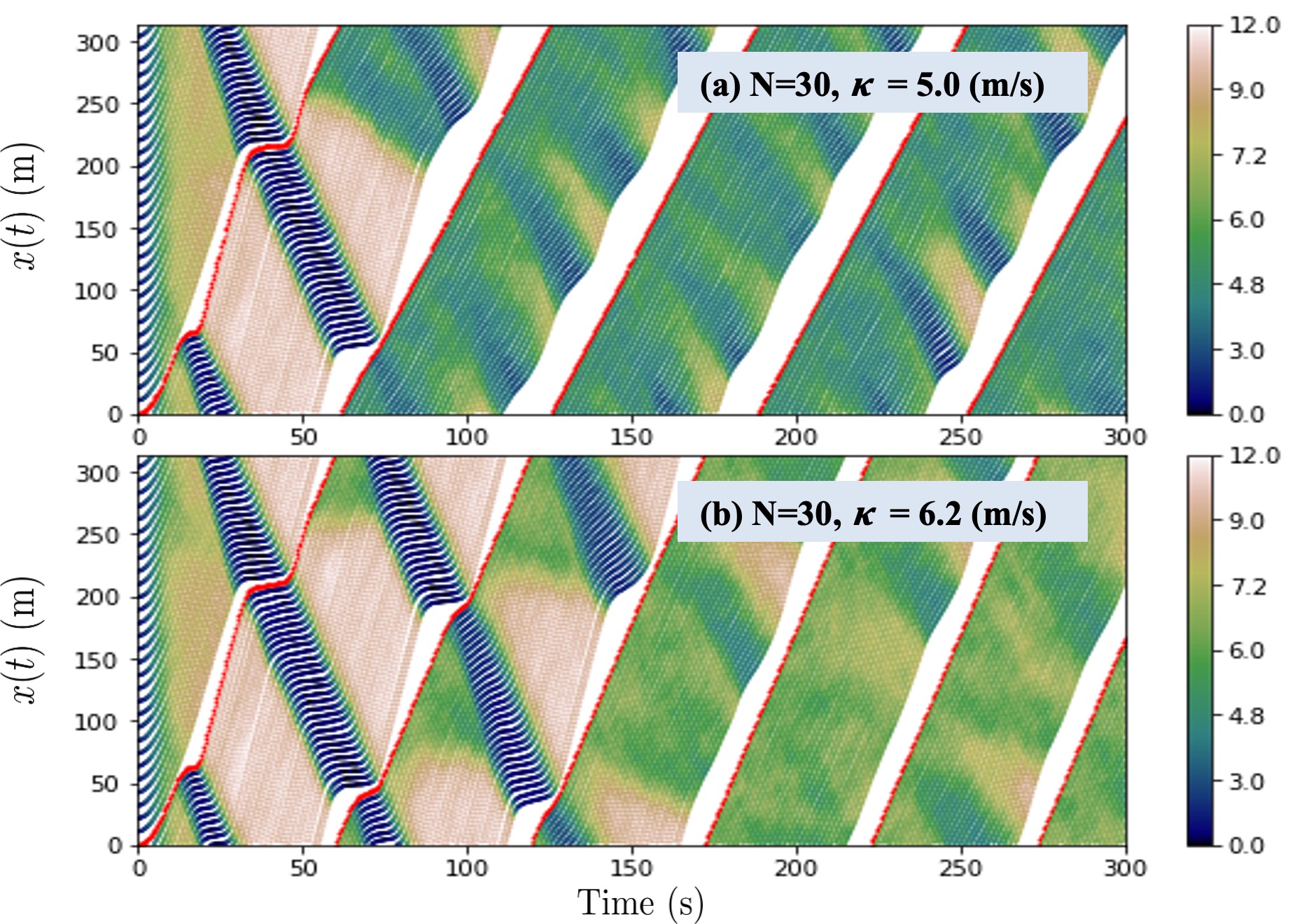}
	\caption{Simulation results with one CAV at $N=30$ in the free-flow phase: Vehicle trajectories become rougher when the value of $\kappa$ becomes lower: (a) $\kappa=5.0$ (m/s); (b) $\kappa=6.2$ (m/s). The red line depicts the trajectory of vehicle \#1.  The color legend represents instantaneous speed of the vehicle in m/s. }
	\label{trajectory_N30_slower}
\end{figure}

The long-run mean functions of the average-speed and speed-range functions, $\bar{V}(\kappa|N)$ and $\bar{R}(\kappa|N)$, can be derived by fitting smooth curves to the blue dots and orange dots in Fig.(\ref{kappa_scan_N30}), respectively. Once these two long-run mean functions are obtained, we can form the objective function $\Pi(\kappa|N,\omega)=\bar{V}(\kappa|N)-\omega\bar{R}(\kappa|N)$.  Optimizing this function with respect to $\kappa$ for each given value of $\omega$ yields the efficient frontier,  defined as $\bar{V}\big(\kappa^*(N,\omega)|N\big)$ vs $\bar{R}\big(\kappa^*(N,\omega)|N\big)$,  as shown in Fig.\ref{efficient_frontier_N30}. The overall shape of the efficient frontier is intuitive. As expected, increasing $\omega$ reduces speed-range. But it is achieved at the expense of reduced average speed. Given the plotting scale of Fig.\ref{efficient_frontier_N30}(b), the $\omega$-dependence of $\kappa^*(N,\omega)$ is relatively weak, indicating that the tradeoff attitude does not matter too much in practice.  The weak dependence of the efficient frontier on $\omega$ is caused by the fact that the peak in average-speed and the valley in speed-range nearly coincide at around $\kappa\sim 6.3$ m/s (see Fig.\ref{kappa_scan_N30}). Together the peak and valley effectively ``trap'' the optimal ideal speed to its vicinity, so long as $\omega$ is not too large. Only when the value of the tradeoff parameter becomes very high ($\omega\gg 1$) the entire system is pushed abruptly into the extreme point of the suppressed flow phase ($\kappa^*\rightarrow 0$), which is an undesirable outcome.

\begin{figure}[!h]
	\centering
		\includegraphics[width=0.46\textwidth]{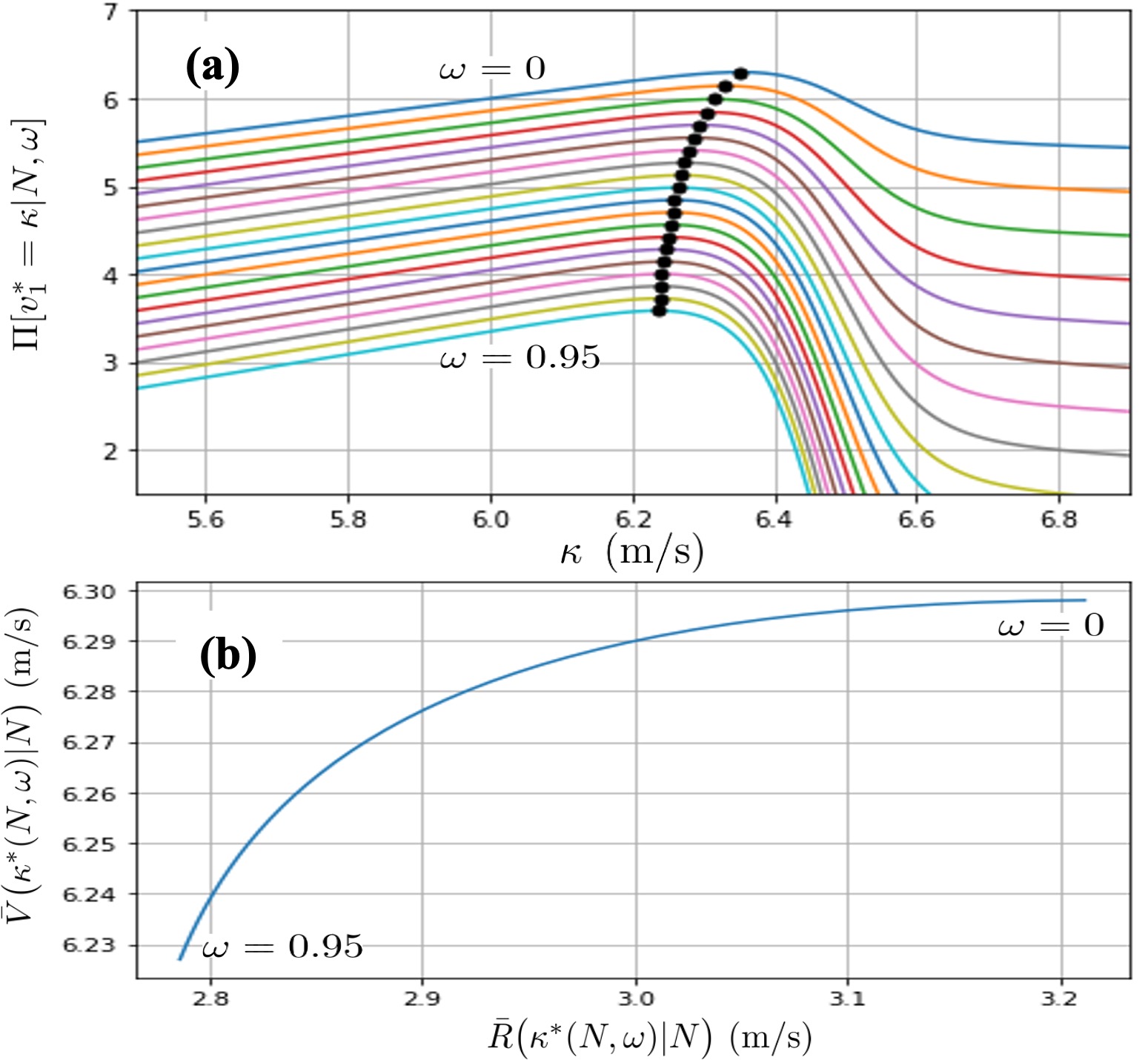}
	\caption{(a) System-level objective function at $N=30$: $\Pi(\kappa|N,\omega)$, with the black dots indicating its maximum points for a given range of $\omega$; (b) Efficient frontier: $\bar{V}\big(\kappa^*(N,\omega)|N\big)$ vs $\bar{R}\big(\kappa^*(N,\omega)|N\big)$, as $\omega$ is varied from $0$ to $0.95$. }
	\label{efficient_frontier_N30}
\end{figure}

\section{\bf The Cases of Multiple CAVs}\label{Multi-CAV}
We now consider the cases when multiple CAVs are involved,  with each CAV controlled independently by the traffic control system.  For simplicity we assume that CAVs are evenly inserted into the traffic system (e.g. $i=1$ and $i=1+N/2$ for the case of 2-CAV) and subject to the same control policy. To get a feeling on how the concentration of CAVs could impact the optimal ideal speed and traffic outcome, we plot in Fig.\ref{kappa_star_vs_N} $\kappa^*(N,\omega=1)$ for cases with one, two and all CAVs (labeled by 1-CAV, 2-CAV and N-CAV, respectively). Note that $\kappa^*(N,\omega)$ nearly  equals the average speed of the regulated traffic. As a comparison, the average-speed of the case without any CAVs is also shown in the same figure (gray line).  We observe that all these curves have relatively mild dependences on $N$, $\omega$, and the concentration of CAVs. This implies some degree of implementation flexibility: even when the optimal ideal speed for CAVs are enforced approximately, the controlled traffic will not suffer too much from noticeable degradation in our improvements to efficiency/smoothness.

\begin{figure}[!h]
	\centering
	\includegraphics[width=0.44\textwidth]{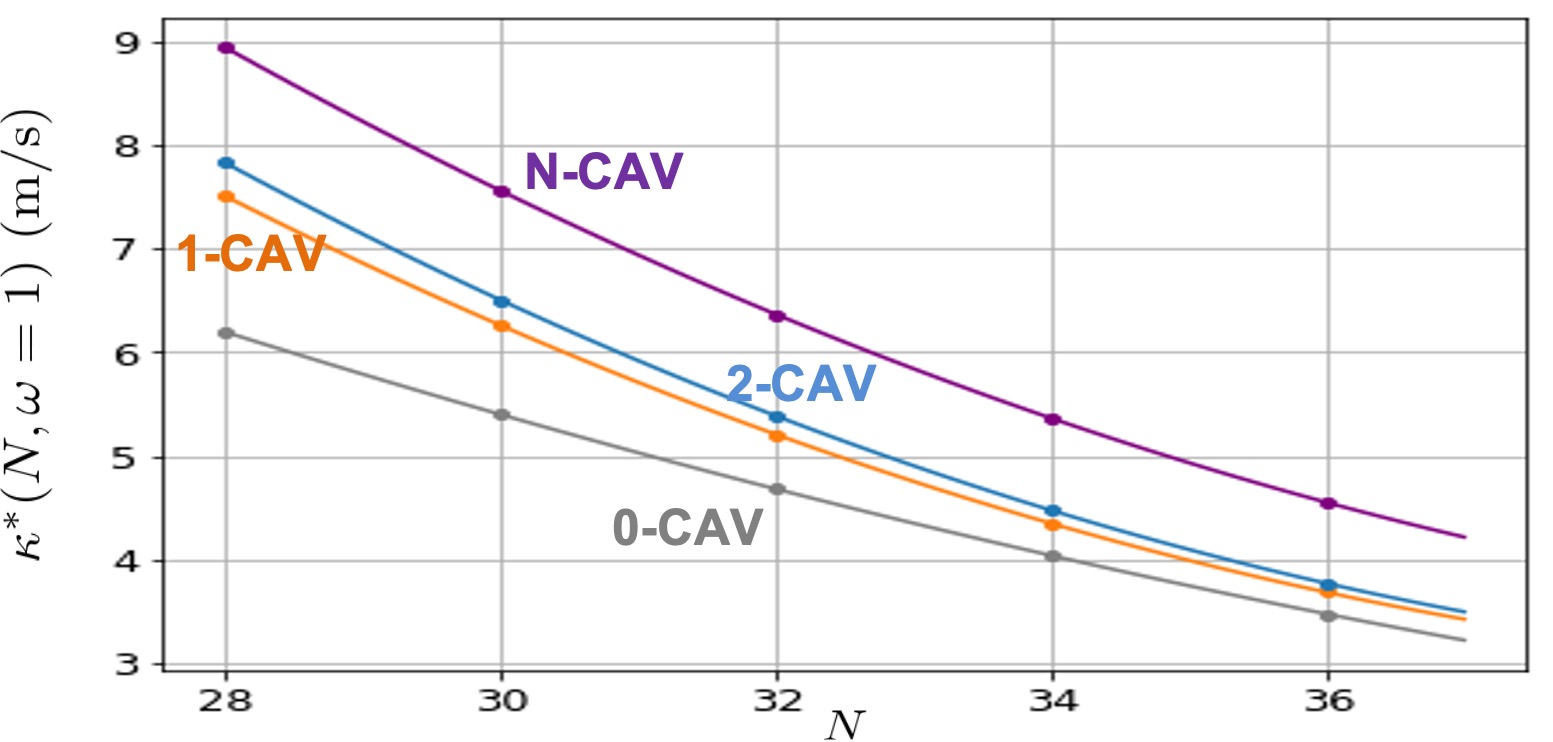}
	\caption{Optimal ideal speed of the system-controlled CAVs: $\kappa^*(N,\omega=1)$.  For comparison purposes, the average speed of the traffic system with all vehicles being human-driven is depicted by the gray line.}
	\label{kappa_star_vs_N}
\end{figure}

Fig.\ref{benefit_plot} shows the quantitative improvements of the average-speed and speed-range as functions of $N$ and $\omega$ relative to those in the uncontrolled jammed flow phase.  Note that these are Pareto improvements, because both $\bar{V}$ and $\bar{R}$ are better in the traffic modulated by a CAV. When only one CAV is inserted optimally, the average-speed is boosted by about $5.6\%$ ($N=36$) to $20.5\%$ ($N=28$), while the speed-range is reduced by 54.1\% ($N=36$) to 81.7\% ($N=28$). The improvement becomes even more pronounced when two or more CAVs are inserted.  If we convert $N$ dependence to vehicle density dependence,  results of 0-CAV and N-CAV in Fig.\ref{kappa_star_vs_N} and Fig.\ref{benefit_plot} become directly comparable to Fig.9 and Fig.10 from \cite{Bifurcation}.  We see that driver heterogeneity and state evolution noise do not significantly alter the ``clean'' version of the corresponding curves derived using bifurcation analysis in \cite{Bifurcation}.

\begin{figure}[!h]
	\centering
	\includegraphics[width=0.47\textwidth]{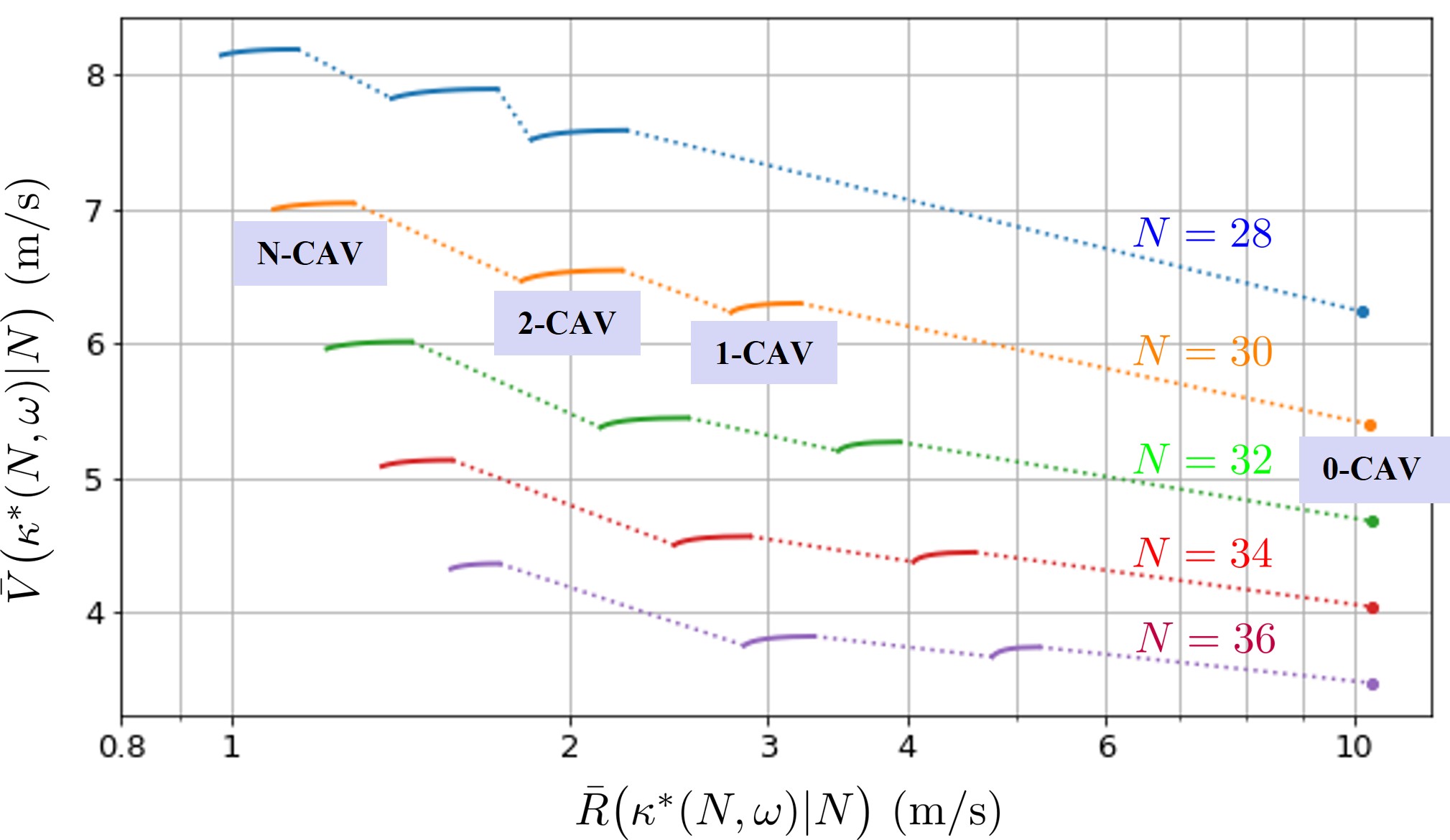}
	\caption{Pareto improvements in long-run mean of average-speed and speed-range relative to the uncontrolled jammed-flow phase (dots on the right). The full lines are the efficient frontiers with one, two and all CAVs at each value of $N$. Note that the horizontal scale is logarithmic.}
	\label{benefit_plot}
\end{figure}

To understand why more CAVs is better than fewer CAVs, we plot simulated vehicle trajectories at $N=30$ with a varied number of CAVs in Fig.\ref{trajectory_N30_CAV}. One immediately recognizes that there are much higher residual waves with 1-CAV than the case with 2-CAV., while N-CAV has least residual waves.  To generate stop-and-go waves, these residual waves require a build-up process: the longer the queue the easier for the residual waves to cumulate. Of course, these are recognized as the precursors to the original waves when there is no CAV present.
\begin{figure}[!h]
	\centering
		\includegraphics[width=0.44\textwidth]{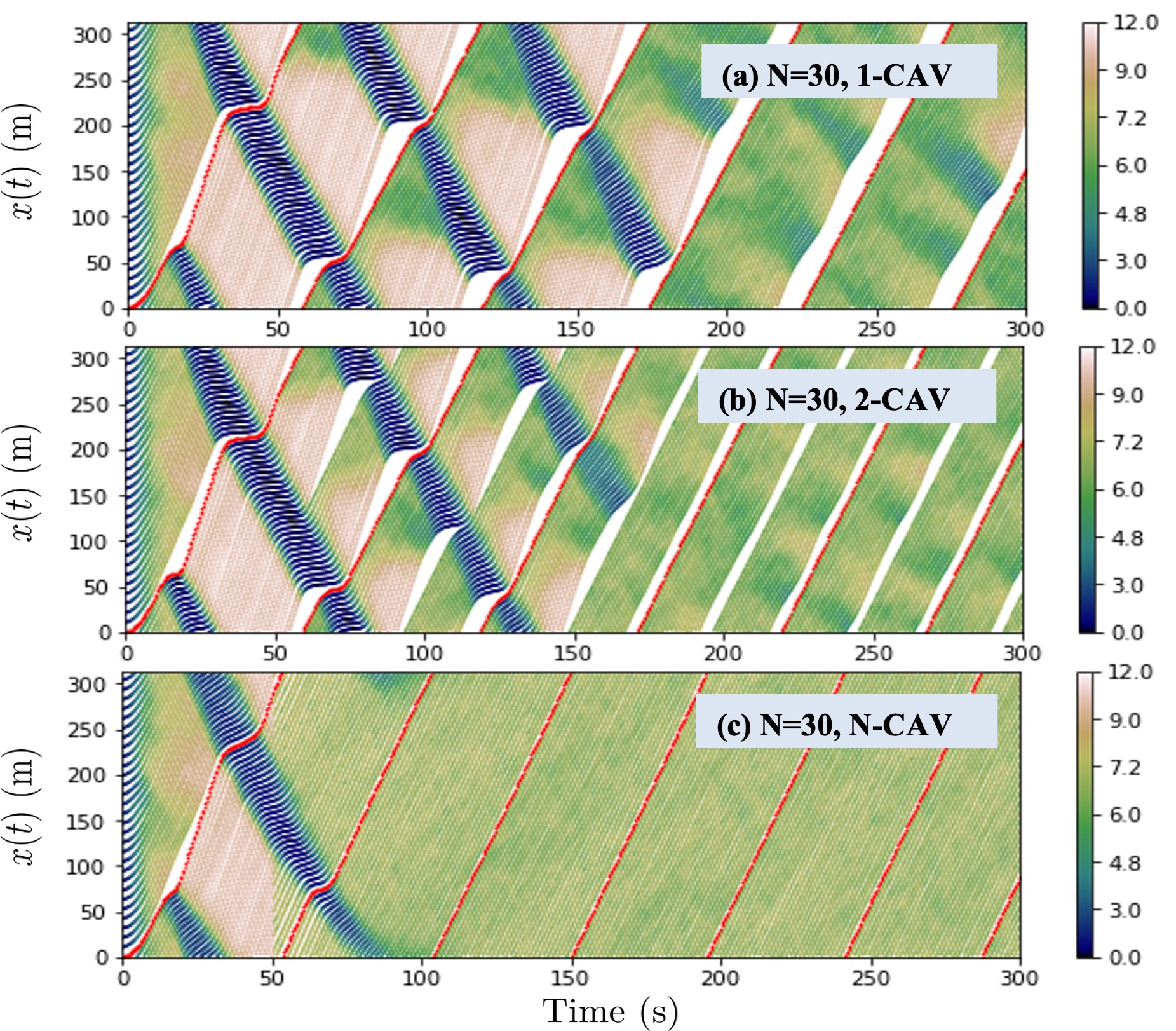}
	\caption{Vehicle trajectories are further smoothened when more CAVs are optimally inserted: (a) $N=30$ with 1-CAV; (b) $N=30$ with 2-CAV; (c) $N=30$ with $N$-CAV.  The red line depicts the trajectory of vehicle \#1.  The color legend represents instantaneous speed of the vehicle in m/s. }
	\label{trajectory_N30_CAV}
\end{figure}

To further elucidate the nature of $\kappa^*(N,\omega)$ theoretically,  it is instructive to examine the case of N-CAV more closely.  Following the same simulation procedure, we can re-calculate $\kappa^*(N,\omega)$ with driver heterogeneity and evolution noise switched off.  It turns out that the ``clean'' version of $\kappa^*(N,\omega)$, plotted as a function of vehicle density ($\rho\equiv N/C$), coincides with the 2D bifurcation curve labeled $\rho^*_\text{GS1}(v^*)$ in Fig.9 from \cite{Bifurcation}.  Despite the use of different techniques to derive them, optimizing a system-level objective function for $\kappa^*(N,\omega)$ versus bifurcation analysis for $\rho^*_\text{GS1}(v^*)$, this coincidence immediately implies that the optimal control policies displayed in Fig.\ref{kappa_star_vs_N} have a common underlying origin in bifurcation, though distorted by driver heterogeneity and evolution noise. It further illustrates that a properly defined system-level objective can effectively detect abrupt transition of the traffic dynamics without technical restrictions required by rigorous bifurcation analysis.

Finally, our findings above indicate that the proposed CAV-modulated traffic control system does not fully remove the instability of the traffic system, only reduces its extent. If the penetration of system-controlled CAVs is sufficiently low, long consecutive strings of human-driven vehicles will be common. Subsequently, stop-and-go waves will continue to form spontaneously. However, as the penetration of system-controlled CAVs increases, there will be fewer human-driven vehicles in a consecutive line. This in turn implies that stop-and-go waves will have less room to grow into something substantial in magnitude, and will not persist for as long a period of time. With such a practical perspective, removing intrinsic traffic instability is no longer the most pertinent question. What really matters is whether we can reduce traffic jams, by improving throughput and smoothness of the controlled traffic simultaneously. However, the tendency for oscillatory waves is still there. On the other hand, when all the ideal speeds are constrained from the above,  such as imposing a speed advisory, the stop-and-go wave can no longer be asymptotically stable, as shown in the recent 2D-bifurcation study \cite{Bifurcation}. Only then will the oscillatory waves be truly tamed.

\section{\bf Possible Realization of the Traffic Control System}\label{System_realization}
The simplicity of the proposed mechanism implies that the system controlled CAV only needs to act like a smart ACC with a system provided ideal cruising speed, which in turn is adapted to the local vehicle-density and the CAV penetration. Consequently, we can envision the following three possibilities to implement the mechanism in practice, assuming that the traffic control system can monitor the local vehicle-density and CAV penetration.

\subsubsection{CAV modulated traffic system} The traffic control system can directly control enough CAVs in traffic. This is exactly what the original design implies. However, system-controlled CAVs may not be immediately available in the short run.
\subsubsection{Cruising speed recommendation system for ACCs} There are enough ACC-ready vehicles in the traffic who are willing to execute the recommended cruising speed from the traffic control system, such as through an electronic message board. The obvious advantage of this implementation is that the availability of ACCs is already becoming prevalent. On the other hand, this indirect method needs to rely on unproven willingness for at least some ACC drivers to follow the system recommendation. Note that, even though it is apparently similar to usual VSA schemes, our decentralized proposal is applied only to ACC-ready vehicles rather than all vehicles at the scene.
\subsubsection{Command-and-control system via VSA} The traffic control system enforces a variable speed advisory that is adapted to the vehicle density of the traffic. The recommended speed effectively puts an upper bound on the ideal speed for all vehicles in the traffic, equivalent to the situation of N-CAVs.\footnote{In practice, all vehicles can be human driven. We only need to assume that the percentage of drivers who are willing to adhere to the VSA is substantial. Additionally, a lower speed limit can be imposed in order to stay away from the suppressed-flow regime per Fig.\ref{kappa_scan_N30}.} In fact, this is not only the simplest way to implement the traffic control, it is also when improvements for both traffic efficiency and smoothness reach their highest potential, as indicated by Fig.\ref{benefit_plot}.

It is important to point out that our proposed mechanism requires only coordination between CAVs (or ACCs) and the roadside infrastructure of the traffic control control system. No explicit vehicle-to-vehicle communications are necessary,  in contrast with the case of CACC systems. Furthermore, the traffic monitoring task of the traffic control system is at ``macro-level'', rather than at micro-level. The latter would have been much more technically demanding. In the decentralized implementation, monitoring the density of ACCs who are willing to following the recommendation may not be totally straightforward. However, given the relatively mild dependence on incremental CAVs in Fig.\ref{benefit_plot}, some inaccuracy in monitoring ACC density may not lead to too much degradation from optimality. 

\section{\bf Summary and Outlook}
In this paper we first demonstrated that the recently proposed human driving behavior model, with its model parameters quantitatively calibrated using vehicle trajectory data from the Sugiyama experiment, can generalize well to the Tadaki experiment and reproduce nearly all the important traffic characteristics observed in a wide range of vehicle density.  Taking advantage of this fact, we then applied computational mechanism design to systematically search for a very simple yet effective traffic control algorithm to optimally tame the stop-and-go waves.  It was shown that the improvements for traffic efficiency and smoothness can be substantial. Although the discussion directly involves system-controlled CAVs, we have also proposed how to apply our findings to a real-world setting without specialized vehicles, {\em indirectly} implementing our algorithm using a VSA.

Admittedly, with a single-lane and closed-ring geometry, the setting of the Tadaki experiment is still somewhat artificial. One of the next steps is to investigate whether the same idea can work in real highway settings. As illustrated in Fig.\ref{US101_waves} the NGSIM data show similar spontaneous formation of stop-and-go waves in a multi-lane and open geometry setting, with major characteristics not being too far from those observed in the Tadaki experiment. To tackle this realistic problem, we first need to build a human behavioral model for highway driving and calibrate it against naturalistic traffic trajectories, generalizing what had been done in \cite{CalibrationPaper}. Exploratory studies along this line had showed encouraging results \cite{HighwayCalibration}. Once it is accomplished, the same computational mechanism design demonstrated here can then be applied by simulating a CAV-modulated highway traffic behavior game, using the calibrated agents as human drivers. It will be interesting to find out whether the improvement will also be Pareto, as was the case we have shown in this paper. Of course, our approach is very general, not just limited to the context of taming waves. Our simulation-based framework with pre-calibrated human driving agents can also be applied to many other settings, ranging from traffic signal systems to newly proposed transportation systems that exhibit pertinent collective behaviors of the underlying traffic systems emerging from the micro level.

\bibliographystyle{IEEEtran}
\bibliography{refs}
	
\end{document}